\documentclass{appolb}
\usepackage{epsfig}
\usepackage{wrapfig}
\newcommand{\lam}{\mbox{$\rm \Lambda ^0$}}
\newcommand{\alam}{\mbox{$\rm \bar \Lambda ^0$}} 

\begin{document}
\title{Measurement of $\Lambda^0$  and $\alam$ polarization in $\nu_\mu$ CC 
interactions in NOMAD}
\author{Dmitry V. Naumov for the NOMAD Collaboration}
\maketitle
\begin{abstract}
The $\Lambda^0$ and $\alam$ polarizations
in $\nu_\mu$ charged current interactions
have been measured in the NOMAD experiment.
The event sample (8087 reconstructed $\Lambda^0$'s and 649 $\alam$'s)
is more than an order of magnitude larger
than that of previous bubble chamber experiments, while the
quality of event reconstruction is comparable.
For the $\Lambda^0$ hyperons 
we observe negative polarization along the $W$-boson direction which
is enhanced in the target fragmentation region:
$P_x (x_F < 0) = -0.21 \pm 0.04 \mbox{(stat)} \pm 0.02 \mbox{(sys)}$.
In the current fragmentation region we find
$P_x (x_F > 0) = -0.09 \pm 0.06 \mbox{(stat)} \pm 0.03 \mbox{(sys)}$.
A significant transverse polarization
(in the direction orthogonal to the $\Lambda^0$ production plane)
has been observed for the first time in a neutrino experiment:
$P_y = -0.22 \pm 0.03 \mbox{(stat)} \pm 0.01 \mbox{(sys)}$.
The dependence of the absolute value of $P_y$ 
on the $\Lambda^0$ transverse momentum
with respect to the hadronic jet direction is in qualitative agreement
with the results from unpolarized hadron-hadron experiments.
The polarization vector of $\alam$ hyperons measured for the first time in 
neutrino interactions is found to be consistent with zero.
\end{abstract}

\vspace*{-0.7cm}
\section{Introduction}
It is very interesting to measure both longitudinal and transverse $\lam$
and $\alam$
polarizations in $\nu N$ deep inelastic process due to different physical 
mechanisms related to them.
When I was presenting first time the NOMAD results on $\lam$ polarization
measurement at SPIN2000 Symposium (Osaka, Japan)~\cite{Naumov:Osaka} I 
followed the common belief that a cut on $x_F$ can separate target remnant
from the struck quark fragmentation regions. This assumption leads to 
a possibility to observe the effect of the polarized intrinsic strangeness at
$x_F<0$ (target fragmentation region) and to measure the spin transfer 
coefficient $C_q^{\lam}$ in quark to $\lam$ fragmentation at $x_F>0$.
However, recently it was found that beam energies of all experiments
involved in $\lam$ polarization measurements are too low to disentagle
quark fragmentation region from the target remnant~\cite{EKN} and all
the experiments deal mainly with left over diquark fragmentation. 
Therefore,  the interpretation of the data at $x_F>0$ should be changed.

The transverse polarization of $\Lambda^0$ hyperons has been observed 
for a long time in unpolarized hadron-hadron 
experiments  ~\cite{lambda_hadron}, and was never  observed in 
(anti)neutrino nucleon DIS experiments 
~\cite{lambda_neutrino}. This surprizing feature challenges experimental and
theoretical efforts in this field. It is interesting to note that the 
transverse polarization of $\alam$ hyperons measured in hadron experiments
is consistent with zero.

Let me skip here the details of the reconstruction and identification
procedures of $\lam$ and $\alam$ hyperons and redirect the interested reader
to the original literature~\cite{NOMAD_lambda,DN_thesis}. 
As a result we obtained 8087 (649) reconstructed and
identified $\Lambda^0$ ($\alam$) hyperons with about $4\%$ ($10\%$)
background  contamination in our data samples.
These samples are used for the polarization analyses reported below.
The $\Lambda^0$ ($\alam$) polarization is measured through the {\em 
asymmetry} in the angular distribution of the protons (pions, $\pi^+$) 
in the parity violating decay process $\Lambda^0 \to p \pi^-$ 
($\alam \to \bar p \pi^+$).
In the $\Lambda^0$ ($\alam$) rest frame the decay protons (pions, $\pi^+$)
are distributed as:
$
\frac{1}{N}\frac{d N}{d \Omega} = 
\frac{1}{4\pi}(1+\alpha_\Lambda \mathbf P \cdot \mathbf k),
\label{eq:asymmetry}
$
where $\mathbf P$ is the $\Lambda^0$ ($\alam$) polarization vector,
$\alpha_\Lambda = 0.642 \pm 0.013$~\cite{PDG} is the decay asymmetry parameter 
and $\mathbf k$ is the unit vector along the decay positively charged
track (proton for $\Lambda^0 \to p \pi^-$ and pion for 
$\alam \to \bar p \pi^+$) direction. We define the quantization axes as 
follows:
\begin{itemize}
\item $\mathbf n_x =  \mathbf e_W$, where $\mathbf e_W$ is the reconstructed 
$W$-boson direction;
\item $\mathbf n_y = \mathbf e_W \times \mathbf e_{T} /
|\mathbf e_W \times \mathbf e_{T} |$ axis is orthogonal to the $\Lambda^0$ 
($\alam$) production plane 
\item $\mathbf n_z = \mathbf n_x \times \mathbf n_y$. 
\end{itemize}
\vspace*{-1cm}

\section{Results and discussion}
\vspace*{-0.2cm}

Table ~\ref{tab:general} displays the results for the polarization of 
$\Lambda^0$ hyperons in our sample as a function of $x_F$. We observe
negative longitudinal (``$P_x$'') and transverse (``$P_y$'') polarizations 
of $\Lambda^0$'s which are enhanced
in the target fragmentation region. Note that transverse polarization has 
never been observed before in (anti)neutrino nucleon DIS experiments. 
The polarization vector of $\alam$ hyperons measured at different cuts on 
$x_F$ and $x_{Bj}$ is shown in Tab.~\ref{tab:antilambda-numu}. One can
conclude that this vector is consistent with zero within the statistical
errors. In what follows we present further results for $\lam$ hyperons.

\begin{table}[htb]
\begin{center}
\caption{\label{tab:general} \it 
Dependence of the $\Lambda^0$ polarization on $x_F$ in $\nu_\mu$ CC events 
(statistical errors only).}
\begin{tabular}{||c|c|c|c|c|c||}
\hline
\hline
 & & & \multicolumn{3}{|c||}{$\Lambda^0$ Polarization} \\
\cline{4-6}
Selection & Entries & $<x_F>$ & $P_x$         &  $P_y$         &$P_z$ \\
\hline
\hline
full sample & 8087 & $-0.18$ & $-0.15 \pm 0.03$ & $-0.22 \pm 0.03$ & $-0.04 \pm 0.03$\\
\hline
\hline
$x_F<0$ & 5608 & $-0.36$ & $-0.21 \pm 0.04$ & $-0.26 \pm 0.04$ & $-0.08 \pm 0.04$\\
$x_F>0$ & 2479 & $0.21$ & $-0.09 \pm 0.06$ & $-0.10 \pm 0.06$ & $ 0.02 \pm 0.06$\\
\hline 
\hline
\end{tabular}
\end{center}
\end{table}

\begin{table}[htb]
\begin{center} 
\caption{\it 
$\bar \Lambda$ polarization in $\nu_\mu$ CC events 
(the  error is statistical).}
\begin{tabular}{||c|c|c|c|c||}
\hline
\hline
 & & \multicolumn{3}{|c|}{$\bar \Lambda$ Polarization} \\
\cline{3-5}
Selection  & Entries  &  $P_x$         &  $P_y$         &$P_z$ \\
\hline
\hline
full sample & 649     &$-0.07 \pm 0.12$&$0.09 \pm 0.13 $&$ 0.10 \pm 0.13$\\
\hline
\hline
$x_F<0$     & 248     &$ 0.23 \pm 0.20$&$0.04 \pm 0.20$&$-0.08 \pm 0.21$\\
$x_F>0$     & 401     &$-0.23 \pm 0.15$&$0.10 \pm 0.17$&$ 0.25 \pm 0.16$\\
\hline
\hline
$x_{Bj}<0.2$& 331    &$-0.12 \pm 0.17$&$ 0.08 \pm 0.18$&$ 0.01 \pm 0.17$\\
$x_{Bj}>0.2$& 318    &$-0.03 \pm 0.17$&$ 0.10 \pm 0.18$&$ 0.20 \pm 0.19$\\
\hline
\hline
\end{tabular}
\label{tab:antilambda-numu}
\end{center} 
\end{table}

\begin{figure}[htb]
\begin{minipage}[r]{0.48\textwidth}
  \centering\epsfig{file=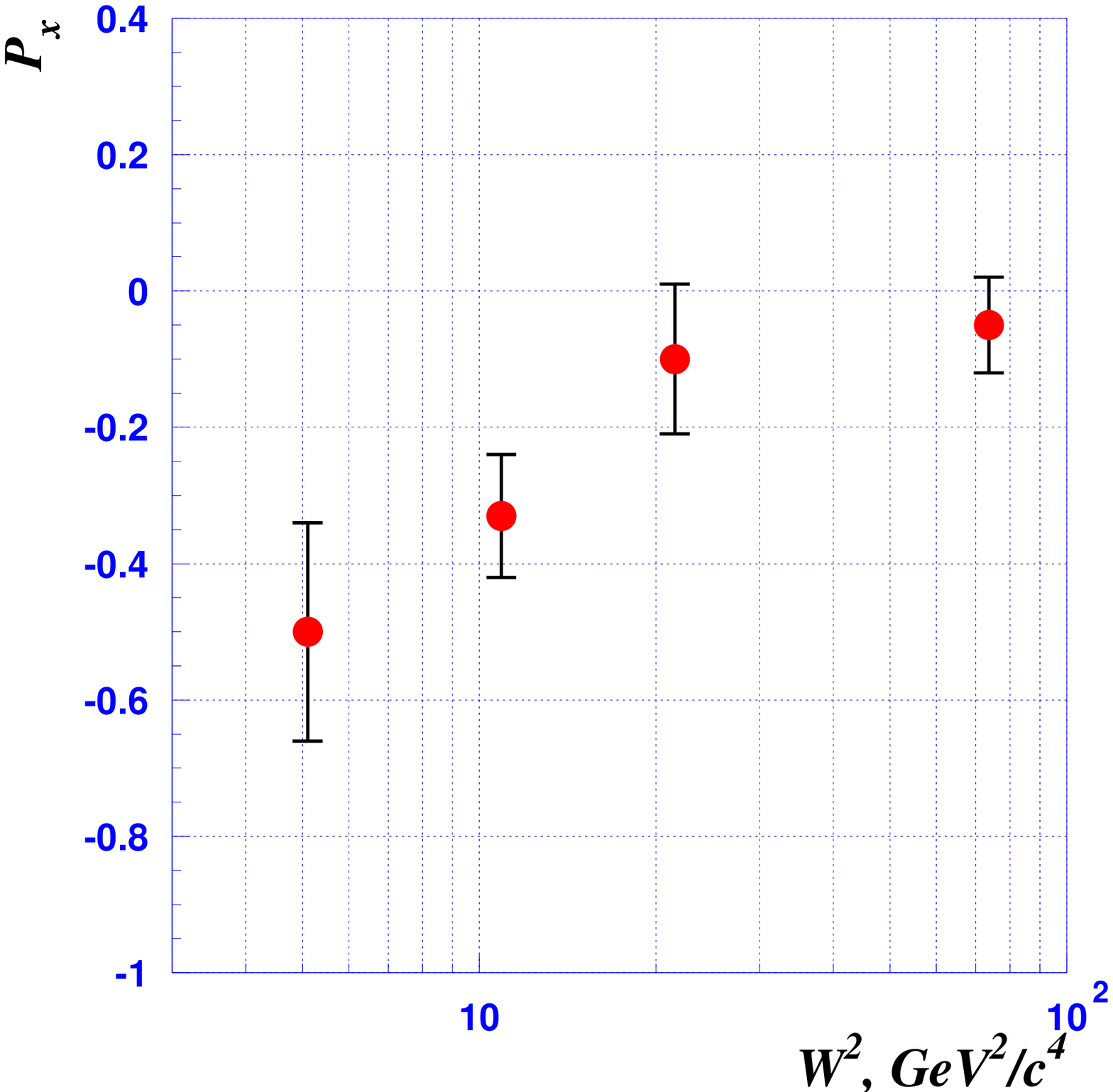,height=4.7cm,width=\textwidth} 
  \caption{\label{fig:long_w2_xfn} Longitudinal polarization of $\Lambda^0$ as a function of $W^2$ for $x_F<0$}
\end{minipage}%
\hspace*{0.02\textwidth}
\begin{minipage}[l]{0.48\textwidth}
  \centering\epsfig{file=EPS/cosy-pt2-xfn.epsi,height=4.7cm,width=\textwidth} 
  \caption{\label{fig:trans_pt_xfn} Transverse polarization of $\Lambda^0$ as a function of $P_T$ for $x_F<0$}
  \end{minipage}
\end{figure}

The dependence of the longitudinal polarization of $\Lambda^0$ on $W^2$ at 
$x_F<0$ is shown in Fig.~\ref{fig:long_w2_xfn}. 
Large negative $P_x$ is observed at small $W^2$, while at 
larger $W^2$ the longitudinal polarization vanishes. 
We have performed a study of the dependence of the transverse polarization on  
the $\Lambda^0$ transverse momentum with respect to the jet direction  ($p_T$) in $x_F<0$ region and found it to be in qualitative 
agreement (both sign and shape) with that found in 
unpolarized hadron-hadron collisions ~\cite{lambda_hadron}. 
Also, we observed no dependence
of $P_y$ on $W^2$. These features make possible to conclude that the origin 
of the transverse polarization is in the fragmentation process.

\begin{figure}[htb]
\begin{center}
\epsfig{file=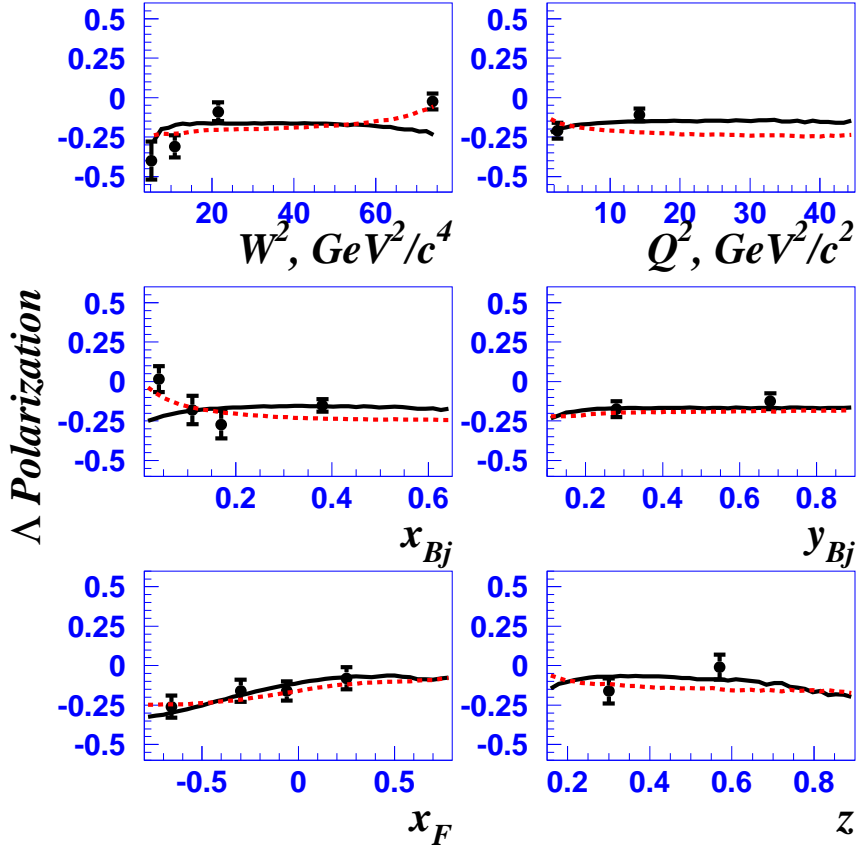,width=\linewidth} 
\caption{\label{fig:numuxcc_6} The predictions of model A - solid line 
and model B - dashed line, for the polarization
of $\Lambda$ hyperons produced in $\nu_\mu$ charged-current DIS
interactions off nuclei as functions of $W^2$, $Q^2$, $x_{Bj}$,
$y_{Bj}$, $x_F$ and $z$ (at $x_F>0$)~\cite{EKN}.}
\end{center}
\end{figure}

Imposing a cut on the sum of charges of the all outgoing tracks 
from the primary vertex ($Q_{tot}$) of the event we can study 
$\Lambda^0$ polarization on different 
target nucleons. We select $\nu_\mu p$ ($\nu_\mu n$)-like events requiring $Q_{tot} \ge 1$  ($Q_{tot} \le 0$)
with purity of the selection 76\% (85\%).
The results are summarized in Table.~\ref{tab:target}. 
There is a strong dependence
of the polarization vector on the type of the target nucleon. 

A comparison of our data to theoretical calculations ~\cite{EKN} 
is presented in Fig.~\ref{fig:numuxcc_6}. The calculations ~\cite{EKN}
are based on SU(6) plus polarized intrinsic strangeness model for the
diquark fragmentation and it uses SU(6) model to take into account 
spin transfer from quark fragmentation. One can conclude that this model 
nicely describes the NOMAD data on the longitudinal polarization 
though it is not as good in describing the 
target nucleon effects observed by the NOMAD (see Tab.~\ref{tab:numuxcc}).
Still there is no model able to address the transverse polarization measured
by the NOMAD Collaboration.

\begin{table}[htb]
\begin{center}
\caption{\label{tab:target}\it
The dependence of the $\Lambda^0$ polarization on the type of target nucleon.}
\begin{tabular}{||c|c|c|c|c||}
\hline
\hline
 & & \multicolumn{3}{|c||}{$\Lambda^0$ Polarization} \\
\cline{3-5}
Target & Entries & $P_x$         &  $P_y$         &$P_z$ \\
\hline
\hline
``proton'' & 3472 & $-0.26 \pm 0.05$ & $-0.09 \pm 0.05$ & $-0.07 \pm 0.05$\\
$x_F<0$    & 2407 & $-0.29 \pm 0.06$ & $-0.10 \pm 0.06$ & $-0.09 \pm 0.06$\\
$x_F>0$    & 1065 & $-0.23 \pm 0.09$ & $-0.06 \pm 0.09$ & $-0.02 \pm 0.10$\\
\hline 
\hline
``neutron''& 4615 & $-0.09 \pm 0.04$ & $-0.30 \pm 0.04$ & $-0.03 \pm 0.05$\\
$x_F<0$    & 3201 & $-0.16 \pm 0.05$ & $-0.37 \pm 0.05$ & $-0.07 \pm 0.05$\\
$x_F>0$    & 1414 & $ 0.01 \pm 0.08$ & $-0.11 \pm 0.08$ & $ 0.04 \pm 0.09$\\
\hline 
\hline
\end{tabular}
\end{center}
\end{table}

\begin{table}[htb]
  \begin{center}
    \caption{\label{tab:numuxcc}\it 
Dependence of the polarization of $\Lambda$  hyperons produced in 
$\nu_\mu$ CC DIS on the type of target nucleon predicted in ~\cite{EKN}
compared with the NOMAD data.}
    \begin{tabular}{c|c|c|c}
\hline\hline
&\multicolumn{3}{c}{Target nucleon}\\
\cline{2-4}
$P_\Lambda$ (\%) & isoscalar & proton & neutron\\
\cline{2-4}
model A& -17.4& -11.4& -20.2\\
model B& -19.3& -18.1& -19.9\\
NOMAD& -15.0$\pm 3$ & -26.0$\pm 5$ &  -9.0$\pm 4$ \\
\hline\hline
\end{tabular}

  \end{center}
\end{table}


\vspace*{-1cm}

\begin{thebibliography}{99}
\bibitem{Naumov:Osaka} D.~V.~Naumov  [NOMAD Collaboration],
  AIP Conf.\ Proc.\  {\bf 570} (2001) 489, hep-ph/0101325
\bibitem{EKN} J.~Ellis, A.~Kotzinian, D.Naumov, hep-ph/0204206
\bibitem{lambda_hadron} {\it 
see review} J.F\'elix, {\it
Mod. Phys. Lett.}{\bf A14} (1999) 827 
\bibitem{lambda_neutrino} G.T.Jones {\it et al.}, {\it Z. Phys.} {\bf C28} 
              (1985) 23; S.Willocq {\it et al.}, {\it Z. Phys.} {\bf C53} 
	      (1992) 207; D.DeProspo {\it et al.}, {\it Phys. Rev.} {\bf D50} 
	      (1994) 6691; V.Ammosov {\it et al.}, {\it Nucl. Phys.} {\bf B162}
	      (1980) 205; D.Allasia {\it et al.}, {\it Nucl. Phys.} {\bf B224}
	      (1983) 1


\bibitem{NOMAD_lambda} P.Astier et al., [NOMAD Collaboration], {\it 
Nucl. Phys.} {\bf B588} (2000) 3; 
Astier {\it et al.}  [NOMAD Collaboration],
Nucl.\ Phys.\ B {\bf 605}, 3 (2001).
\bibitem{DN_thesis} 
  D.~V.~Naumov, Ph.D. Thesis, JINR, Dubna, Russia, (2001), {\it 
in Russian}. (available upon request).
\bibitem{PDG} Review of Particle Properties, {\it Eur. Phys. J.}
	      {\bf C15} (2000)



\end{thebibliography}
\end{document}